\documentclass[12pt]{article}

\newcommand{\be}{\begin{equation}}
\newcommand{\ee}{\end{equation}}
\newcommand{\bea}{\begin{eqnarray}}
\newcommand{\eea}{\end{eqnarray}}

\newcommand{\bit}{\begin{itemize}}
\newcommand{\eit}{\end{itemize}}

\newcommand{\beqs}{\begin{eqnarray}}
\newcommand{\eeqs}{\end{eqnarray}}
\begin{document}
\bibliographystyle{h-physrev}
{\sf \title{Non-local symmetries for Yang-Mills theories and their massive counterparts in two and three dimensions}
\author{Abhishek Agarwal\footnote{email: abhishek@aps.org} and Ansar Fayyazuddin\footnote{email: ansarf@aps.org} }
\maketitle
\begin{center}
\vspace{-1cm}
{\it American Physical Society \\ 1 Research Rd\\Ridge, NY}
\end{center}

\begin{abstract}
We identify a non-local symmetry for Yang-Mills theories in $1+1$ and $2+1$ spacetime dimensions.  The symmetry mixes a vector current with the gauge field.  The current involved in the symmetry is required to satisfy certain constraints.  The explicit solution for the current obeying these constraints, is obtained in two spacetime dimensions and in the abelian case in three dimensions. We conjecture that the current is generated from a non-local gauge and Lorentz invariant mass term in three dimensions and provide some evidence for it.
 We also posit a conserved current associated with the symmetry generators and derive some of its properties.  In the Abelian case, we compute the symmetry algebra and show that additional symmetry generators have to be included for the algebra to close.  The algebra contains an SO(2,1) subalgebra.  We also comment on the implications of this symmetry for $\mathcal{N}=1$ supersymmetry.
\end{abstract}

\vspace{-16cm}
\begin{flushright}
hep-th/yymmnnn
\end{flushright}

\thispagestyle{empty}

\newpage

\section{Introduction and outline}
We uncover a novel symmetry of Yang-Mills theories in $1+1$ and $2+1$ dimensions. The symmetry presented in the paper involves the exchange of gauge fields with a current, $J_\mu$, obeying the equations:
\be
D^\mu J_\mu = 0, \hspace{.3cm}  D_{[\mu}J_{\nu]} = F_{\mu \nu} \label{fund-constraints}.
\ee
Along with the requirement that the current transforms in the adjoint representation, these equations  define $J_\mu$.  

In the case of two spacetime dimensions, we show that currents following from gauged Wess-Zumino-Witten (WZW) models, which in turn can be regarded as mass-terms for the gluons,  satisfy the above constraints. In the case of three spacetime dimensions, the constraints can be solved in the abelian case by a current derived from a non-local mass term.  Although we are able to show that a current satisfying the constraint given above will generate a symmetry even in the non-abelian case, we cannot yet present an explicit solution to these constraints in three dimensions. However, as we argue in this paper, gauge and Lorentz invariant mass terms may serve as good candidates for generating solutions to these constraints. 

The mass-terms that we consider in this paper can be thought of as non-Abelian completions of the non-local term
\be
S_{Abelian} = \frac{m^2}{2}\int_k A^a_\mu(-k) \left(\eta ^{\mu \nu} - \frac{k^\mu k^\nu}{k^2}\right)A^a_\nu(k),
\ee
which is Lorentz and gauge invariant by itself in the Abelian limit in any number of spacetime dimensions. Since not many Lorentz and gauge invariant mass terms are known for gluonic degrees of freedom, it is worth commenting on the nature of the mass terms considered in this paper. In the case of two spacetime dimensions,  gluons can be made massive by   coupling them to  gauged WZW actions.  Such massive gluons have been studied extensively in the past in the  literature on  1+1 dimensional QCD in the bosonized framework\cite{armoni1,armoni2,grossk}.  As mentioned at the outset,  we show that such mass terms can also be seen to generate currents which can be used to define a novel non-local symmetry of two dimensional Yang-Mills theories.

Gluonic mass-terms are much harder to formulate in three spacetime dimensions, however a few examples of candidate mass terms are indeed known in three dimensions. For instance, a Lorentz and gauge invariant generalization of the gauged WZW action to three dimensions  was provided by Alexanian and Nair (AN), who also showed that the AN mass-term is  generated non-perturbatively  in the effective action of pure Yang-Mills in three dimensions upon a gauge invariant evaluation of the re-summed gluon propagator\cite{AN}. This fact  can be regarded as  a manifestly Lorentz covariant way of understanding the appearance of a mass-gap in the spectrum of Yang-Mills in $D=2+1$. A deeper and exact understanding of the mass-gap can be obtained in the gauge invariant Hamiltonian framework of Kim, Karabali and Nair\cite{KKN-Papers} where the mass-gap is understood as a consequence of the finiteness of the volume of the gauge invariant configuration space of the purely gluonic theory. The volume measure on the configuration space induces a mass-term in the Hamiltonian whose covariantization has been shown to generate two independent choices of gauge and Lorentz invariant mass terms in three dimensions\cite{KKN-Mass} - one of the choices being the AN mass term alluded to earlier.  Without being wedded to any particular choice of a mass term, we show that if the current following from a mass-term satisfies (\ref{fund-constraints}), then that current will also separately generate non-local symmetries for the Yang-Mills action as well as for the mass term from which the current follows.

This paper is organized as follows. Section 2 is devoted to the study of the non-local `current-field' symmetry in $D=1+1$. In the following section, we show how such a non-local symmetry can be readily generalized to three dimensions. However, unlike in the case of two spacetime dimensions we will not be able to present a concrete formula for a mass-term - except in the Abelian case - whose current satisfies the constraints necessary for it to generate the non-local symmetry. We do, however, discuss and introduce the AN mass-term in $2+1$ dimensions in a self-contained manner and present some plausibility arguments in favor of it generating a current that satisfies the necessary constraints. 
 In the same section, we also comment on some implications of this potential symmetry for supersymmetry.  In section 4, we propose a conserved current associated with the symmetry and identify some of its properties.  In section 5, we compute the algebra of symmetry generating charges in the Abelian case. In particular, we show that  the iterated application of the symmetry transformations generate new non-local symmetries, and the algebra of symmetry generators has a closed $SO(2,1)$ sub-algebra (which is distinct from the $SO(2,1)$ Lorentz symmetry). 

\section{$YM_{1+1}$, Massive gluons and a Current-Field Symmetry}\label{currentfield2}
A familiar context for studying mass terms for gluons in $D=1+1$ is provided by   two dimensional $QCD$ with massless fermions \cite{armoni1, armoni2,grossk} where  the mass-term can be thought of as the effect of integrating out the fermions. One can equivalently regard the mass-term as the  two dimensional Dirac determinant. This mass-term has the interesting effect of  rendering the $1+1$ dimensional strong force short ranged, leading to screening - as opposed to confinement - in 2D QCD with (strictly) massless fermions\cite{grossk}.  The mass-term for gluons generated upon integrating out massless fermions  is nothing but the Dirac determinant, a regularized and gauge invariant expression for which  can be provided in terms of a  gauged Wess-Zumino-Witten (WZW) functional\footnote{A very lucid summary of the connection of the Dirac determinat to the WZW model can be found in \cite{vpn-book}-chapter-17.}. To be explicit, we introduce the complex coordinates $z = x^i - ix^2,$ $ \bar{z} = x^i + ix^2$. Working with Euclidean signature, one introduces the complex matrix $M$ to parametrize the gauge potentials as 
\be
A = -\partial M M^{-1}
\ee  
Under local gauge transformations generated by the unitary matrix $U$, $M \rightarrow UM$.  Thus the associated hermitian matrix $H = M^\dagger M$ is gauge invariant. The Dirac determinant can be related to a WZW functional with $H$ as the dynamical variable by the following relation \cite{vpn-book}:
\be
\det (D\bar{D})_R = \det (\partial \bar{\partial}) e^{\mathcal{A}_R S_{WZW}[H]}
\ee
$R$ stands for the representation of the massless fermions under the gauge group (generated by $t^a$). 
\be
\mathcal{A}_R = tr(t^a t^b)_R/tr(t^at^b)_{f},
\ee
where $f$ denotes the fundamental representation\footnote{In our conventions $A = -it^a A^a$, tr$(t^at^b) = \frac{1}{2}\delta^{ab}$.  }. 
\be
S_{WZW}[H] = \frac{1}{2\pi}\int_{\mathcal{R}^2}tr(\partial H\bar \partial H^{-1}) + \frac{i}{12\pi}\int \epsilon^{ijk}tr(H^{-1}\partial_i H H^{-1}\partial_j H H^{-1}\partial_k H).
\ee
The first integral is over the complex plane, whereas the second one involves an extension of the plane to a three dimensional space, of which the plane is a boundary\cite{vpn-book}. Using the Polyakov-Wiegmann identity $S_{WZW}$ can be written also as:
\be
S_{WZW}[H] = S_{WZW}[M] + S_{WZW}[H] -\frac{1}{2\pi}\int A^a\bar{A}^a 
\ee
The variation of the WZW functional can be computed using the relations \cite{vpn-book}
\be
\delta  S_{WZW}[M]  = \frac{1}{2\pi}\int \bar{a}^a\delta A^a, \hspace{.3cm} \delta  S_{WZW}[M^\dagger]  = \frac{1}{2\pi}\int a^a\delta \bar{A}^a\label{variation}
\ee 
where:
\be
 a = (M^\dagger )^{-1} \partial M^\dagger, \hspace{.3cm} \bar a = -\bar \partial M M^{-1}
\ee
Thus: 
\be
\delta S_{WZW}[H] = -\frac{1}{2\pi}\int \left(J^a\delta  \bar{A}^a + \bar{J}^a \delta  A^a \right)
\ee 
with $J^a =  (A-a)^a, \bar{J}^a = (\bar{A} - \bar{a})^a$. It can be seen from their definitions that the auxiliary fields $a, \bar{a}$ satisfy the `eikonal' equations:
\be
D\bar{a}^a = \bar{\partial}A, \hspace{.3cm} \bar{D}a^a = \partial \bar{A}^a.\label{eikonal}
\ee
The currents, $J, \bar{J}$  are associated with  a holomorphic (and anti-holomorphic) symmetry of the WZW functional. For instance, $S_{WZW}[H]$ is invariant under the infinitesimal left (right) translation of $H$ by a holomorphioc (anti-holomorphic) function: $\delta H = \theta (z) H$. The current associated with this symmetry is $-\partial H H^{-1}$ which can be related to $J$ as $ -\partial H H^{-1} = M^\dagger J M^{\dagger -1}$. There is a second symmetry generated by the same currents. It is readily seen using (\ref{variation}) that:
\be
\delta A = i J \epsilon, \hspace{.3cm} \delta \bar{A} = -i\bar{J} \epsilon\label{2dsymm}
\ee  
where $\epsilon $ is real, is a symmetry of $S_{WZW}$. What is surprising, but nevertheless true, is that (\ref{2dsymm}) is also a symmetry of the pure Yang-Mills action in $D=1+1$.  To verify this claim, we note that\footnote{In the complex notation $F = D\bar{A} - \bar{\partial} A$. }:
\be
\delta \frac{1}{4} \int F^a_{\mu \nu} F^a_{\mu \nu} = -2\int F^a \delta F^a = -2\int F^a(D\delta \bar{A} - \bar{D} \delta A)^a = 2i\epsilon \int F^a(D\bar{J} + \bar{D}J)^a\label{2dinvariance}
\ee
The expression in the parentheses in the last term vanishes due to the `eikonal' equations (\ref{eikonal}). Thus (\ref{2dsymm}) is a symmetry of both pure Yang-Mills theory as well as of  the $S_{WZW}$ `mass-term' that one can add to it in two spacetime dimensions. More generally it is a symmetry of each of the two terms of
\be
S^{1+1}_m = -2\int F^a F^a  - m^2 S_{WZW}[H] \label{massive2dym},
\ee 
which can also be regarded as an effective action for 2D QCD with massless fermions. In the above equation $m$ is an arbitrary number with the dimensions of mass, but in the context of massless fermions coupled to Yang-Mills theory in two dimensions it will be a specific function of the Yang-Mills coupling, flavor symmetry group of the fermions and the representation of the color group that the fermions transform under. 
\section{Current-Field Symmetry for $YM_{2+1}$ and Mass-Terms for Gluons}
We now study the question of whether the non-local symmetry of two dimensional Yang-Mills theory might have a `lift' to three spacetime dimensions. Phrased differently, we will explore the possibility that there is a symmetry involving a Lorentz covariant current $J_\mu$ in analogy to the way that (\ref{2dsymm}) was a symmetry of  $YM_{1+1}$

In what follows, we will assume that we have a current, $J_\mu$, which transforms in the adjoint representation of the gauge group and satisfies the identities:
\bea
&&D_\mu J_\nu - D_\nu J_\mu = F_{\mu\nu} \label{def}\\
&&D_\mu J^\mu = 0 \label{inv}.
\eea
While we do not need to derive this current from an action, we will explore the consequences of a mass term associated with the current in the following sense.  Suppose that we start with an action:
\be
S= \int d^3x(\frac{1}{4g^2}F_{\mu\nu}^aF^{a\mu\nu}) + \frac{m^2}{g^2}\Gamma(A)
\ee
where $\Gamma$ is an as-yet-unknown mass term.   We relate the current,  $J^a_\mu$, to the variation of the mass term through:
\be
\Gamma(A_\mu+\delta A_\mu) = \Gamma (A_\mu) + \int d^3x J^{a\mu}(x)\delta A^a_\mu(x) + \cdots.
\ee
We first note that in the abelian theory, taking $\Gamma$ to be of the form mentioned in the introduction,
\be
S = \int d^3x(\frac{1}{4g^2}F_{\mu\nu}F^{\mu\nu})  + \frac{m^2}{2g^2}\int d^3x A_\mu\left(\delta^{\mu \nu} - \partial^\mu\frac{1}{\partial_\alpha \partial^\alpha}\partial^\nu\right)A_\nu\label{s}
\ee
yields the current  $J_\mu  = A_\mu -  \partial_\mu\frac{1}{\partial_\alpha \partial^\alpha}\partial^\nu A_\nu$, which indeed satisfies the defining relations (\ref{def}) and (\ref{inv}).  
In the non-abelian case, we are not able to find an explicit formula for the current or an associated mass-term.  Nevertheless, we will explore the consequences of this current being derived from a putative mass term through its variation in the above sense.

Our central result is that the transformation:
\be
\delta A_\mu  = \epsilon_{\mu \nu \rho}\delta \theta ^\nu J^\rho \label{symm},
\ee
where $\delta\theta^\mu$ is a vector of infinitesimal parameters, is a symmetry of both the mass term and the usual Yang-Mills action.

It is straightforward to show that the putative mass term is invariant under the above infinitesimal transformation:
\bea
\delta\Gamma &=& \int d^3x J^{a\mu}(x)\delta A^a_\mu(x)  \\
&=& \int d^3x J^{a\mu}(x)\epsilon_{\mu\nu\rho} \delta\theta^\nu J^{a\rho}(x) \\
&=& 0
\eea

The Yang-Mills action is also invariant, but one has to work a little harder to show this.  Since under an arbitrary variation, the field strength transforms as:
\be
\delta F_{\mu\nu} = D_\mu \delta A_\nu - D_\nu \delta A_\mu
\ee
the variation of $F^2$ under  (\ref{symm}) is:
\bea
&&\!\!\!\!\!\!\!\!\!\!F^{a\mu\nu}\delta F_{\mu\nu}^a = 2F^{a\mu\nu}(D_\mu \delta A_\nu)^a \nonumber  \\
&=&  F^{a\mu\nu} \epsilon_{\nu\rho\sigma} \delta\theta^\rho (D_\mu J^{\sigma})^a \\
&=&2\delta\theta^0(-F^{a01}(D_0J^2)^a + F^{a02}(D_0J^1)^a - F^{a21}(D_2J^2)^a + F^{a12}(D_1J^1)^a) \nonumber \\
&+&2\delta\theta^1(F^{a10}(D_1J^2)^a - F^{a12}(D_1J^0)^a + F^{a20}(D_2J^2)^a - F^{a02}(D_0J^0)^a) \nonumber \\
&+&2\delta\theta^2(F^{a01}(D_0J^0)^a - F^{a10}(D_1J^1)^a + F^{a21}(D_2J^0)^a - F^{a20}(D_2J^1)^a) \nonumber
\eea
Using (\ref{inv}), the above reduces to:
\bea
F^{a\mu\nu}\delta F_{\mu\nu}^a &=& -\delta\theta^\sigma\epsilon_{\mu\nu\rho}F^{a\mu\nu}(D_\sigma J^\rho)^a \nonumber \\
&=& -\delta\theta^\sigma\epsilon^{\mu\nu\rho}F_{\mu\nu}^a((D_\sigma J_\rho)^a - (D_\rho J_\sigma)^a + (D_\rho J_\sigma)^a)
\eea
Aside from raising and lowering indices we have added and subtracted $(D_\rho J_\sigma)^a$ to help simplify the expression.  Using (\ref{def}) the above becomes:
\bea
F^{a\mu\nu}\delta F_{\mu\nu}^a &=& -\delta\theta^\sigma\epsilon^{\mu\nu\rho}F_{\mu\nu}^a(F_{\sigma\rho}^a + (D_\rho J_\sigma)^a) \nonumber \\
&=& -\delta\theta^\sigma\epsilon^{\mu\nu\rho}F_{\mu\nu}^a(D_\rho J_\sigma)^a \nonumber \\
&=& -\partial_\rho(\delta\theta^\sigma\epsilon^{\mu\nu\rho}F_{\mu\nu}^aJ_\sigma^a) +
\delta\theta^\sigma J_\sigma^a\epsilon^{\mu\nu\rho}(D_\rho F_{\mu\nu}^a) \nonumber \\
&=&  -\partial_\rho(\delta\theta^\sigma\epsilon^{\mu\nu\rho}F_{\mu\nu}^aJ_\sigma^a) \label{varact}
\eea
The last equality follows from the Bianchi identity $\epsilon^{\mu\nu\rho}(D_\rho F_{\mu\nu}^a)
 =0$.  While the second equality above follows from the identity:
 \bea
&& \delta\theta^\sigma\epsilon^{\mu\nu\rho}F_{\mu\nu}^aF_{\sigma\rho}^a  \nonumber \\&=&  \delta\theta^0(F_{01}^aF_{02}^a +  F_{20}^aF_{01}^a) +   \delta\theta^1(F_{12}^aF_{10}^a +  F_{01}^aF_{12}^a) +  \delta\theta^2(F_{12}^aF_{20}^a +  F_{20}^aF_{21}^a)  \nonumber \\&=& 0
 \eea
 Thus the variation of the Lagrangian is a total derivative:
 \be
 \delta L = \frac{1}{2}\partial_\rho(\delta\theta^\sigma\epsilon^{\mu\nu\rho}F_{\mu\nu}^aJ_\sigma^a)
 \ee
and the action is invariant.
\subsection{A potential solution to the constraints}
The natural question that arises is whether or not there is an actual solution to the constraints (\ref{def}, \ref{inv}) in three spacetime dimensions.  While we will not be able to answer this question definitively in this paper, we point out a very suggestive mass term due to Alexanian and Nair which might provide a potential solution.

The three dimensional mass-term in question is a particularly elegant `lift' of the two dimensional functional, $S_{WZW}$, to three Euclidean spacetime dimensions constructed in \cite{AN}, which we briefly review below. Key to many physical attributes of $S_{WZW}$ discussed above is the underlying complex structure presented by the two dimensional spacetime. Even though there is no preferred complex structure in three spacetime dimensions, one can construct an infinity of them using the angular variables of an auxiliary sphere ($\Omega$) as parameters.   Concretely, the three vectors
\beqs
&X_0 = (\sin\theta \cos\phi, \sin\theta \sin\phi, \cos\theta)\nonumber\\
&X_1 =   (-\cos\theta \cos\phi, \cos\theta \sin\phi, \sin\theta)\nonumber\\
&X_2 =   (-\sin\theta, \cos\phi,0)
\eeqs
define an orthogonal coordinate system for all values of $\theta $ and $\phi$. The complex  null vectors 
\be
n = X_1 + i X_2, \hspace{.3cm} \bar n = X_1-iX_2
\ee
can be regarded as defining a complex structure for every point on $\Omega$,\footnote{We emphasize that $\Omega$ has nothing to do with the spacetime, which is $R^3$ or $R^{1,2}$.} which can be used to construct quasi-two dimensional complex quantities out of three dimensional $SO(3)$ covariant vectors. For instance, starting with the gauge potentials $A_\mu$, one can form:
\be
A_+ = \frac{1}{2} A_\mu n^\mu, \hspace{.3cm} A_- = \frac{1}{2} A_\mu \bar n^\mu
\ee
Three dimensional$SO(3)$ invariant expressions can be recovered by integrating out the $\Omega$ dependence of the corresponding quantities which are scalars in the quasi two-dimensional sense. For example:
\be
A_\mu A^\mu = \frac{3}{2\pi}\int_\Omega A_+A_-
\ee
Integration over $\Omega $ can be thought of as averaging over the set of all complex structures in three dimensions and $SO(3)$ invariant expressions in three dimensions can - as illustrated above - be generated in a twistorial form by averaging over this infinite family of complex structures.

The covariant analog of $S_{WZW}$ was proposed in \cite{AN} to be 
\[ \Gamma[A]_{AN} = \int dX_0 d\Omega K(A_+,A_-) \label{SM},\]
where the kernel $K$ is given by
\[
K(A_+,A_-) = -\frac{1}{\pi}\int_1\left(tr (A_+(1)A_-(1)) + i\pi I(A_+(1)) + i\pi I(A_-(1))\right).
\]
While
\[ I(A(1)) = i\sum_n \frac{(-1)^n}{n}\int _{2\cdots n}\frac{tr (A(1)\cdots
  A(n))}{\bar{z}_{12}\bar{z}_{23}\cdots \bar{z}_{n1}}
\frac{d^2x_1}{\pi}\cdots \frac{d^2x_n}{\pi}.\label{I}\]

The arguments of $A$ refer to the different `spatial' points $X_\pm$. The transverse coordinate $X_0$ is the same for all
the $A$'s in the above expression for $I$. 

The expression for $K$ is an exact parallel of the two dimensional expression$S_{WZW}$. Continuing with the parallels, the variations of $K$
are given by
\be
\frac{\delta K}{\delta A^a_+} = \frac{1}{2\pi}J^a_- = \frac{1}{2\pi} (A_-  - \mathcal{A}_-)^a, \hspace{.3cm}  \frac{\delta K}{\delta A^a_-} = \frac{1}{2\pi}J^a_+ = \frac{1}{2\pi} (A_+  - \mathcal{A}_+)^a
\ee
where the auxiliary  fields $\mathcal{A}_\pm$ satisfy the equivalents of (\ref{eikonal}):
\be 
D_\pm \mathcal{A}^a_\mp = \partial_\mp A^a_\pm\label{eikonal3}
\ee
The equations above are obviously very suggestive. The current $J_\mu \sim \int _\Omega (J_+\bar n _\mu + J_-n_\mu)$ does satisfy $D_\mu  J^\mu = 0$.  However, it is not clear that after the integration over the angular variables is carried out $D_{[\mu}J_{\nu]} = F_{\mu \nu}$ is satisfied. It may be noted though, that  a plausibility argument for (\ref{eikonal3}) to imply (\ref{def}) was provided in\cite{aa-msusy1}. To really understand whether or not the current derived from the AN mas term satisfies (\ref{def}) one would need to better inderstand the infinite set of non-local interaction vertices contained in the mass-term, which we hope to address elsewhere.
\subsection{Implications for $\mathcal{N}=1$ supersymmetry}
Although our focus in this work is on purely bosonic theories, the current-field symmetry discussed above has important consequences for minimal supersymmetry in three spacetime dimensions. As shown in\cite{aa-msusy1},
the Yang-Mills action deformed by the AN mass term admits an $\mathcal{N}=1$ supersymmetrization. Focusing on the Abelian limit, we point out that the symmetry described above is crucial for the consistency of the supersymmetry transformations given in \cite{aa-msusy1}. In the Abelian case, it is easy to see that:
\be
S_{\mathcal{N}=1} = \frac{1}{g^2}\left(\int \frac{1}{4}F_{\mu \nu}F^{\mu \nu} + m^2\Gamma(A) + \frac{1}{2}\int\bar{\Psi}( \sigma_\mu \partial^\mu + m)\Psi \right)
\ee
is invariant under the supersymmetry transformations\footnote{The Majorana condition in $\bar \Psi = -i\Psi^t\sigma^2$ is implied}:
\bea
\delta_\alpha A_\mu &=& \bar{\alpha}\sigma_\mu \Psi\nonumber \\
\delta_\alpha \Psi &=& -\frac{i}{2}\epsilon_{\mu \nu \rho}F^{\mu \nu}\sigma^\rho \alpha + m J^\mu \sigma_\mu \alpha \label{massivesusy}
\eea
The second term in the fermionic variation represents a mass-deformation of the supersymmetry algebra which the presence of the massive degrees of freedom necessitate. Evaluation of the commutator of the supersymmetry transformations on the gauge field yields:
\be
\delta_{[\rho}\delta_{\omega]}A_\mu = -2F_{\mu\lambda}(\bar \omega \sigma ^\lambda \rho) + 2im\epsilon_{\mu \nu \lambda}J^\mu (\bar \omega \sigma ^\lambda \rho)\label{closure} 
\ee
The first term on the $r.h.s$ has the usual interpretation of a combination of translation and gauge transformation of the gauge field. The second term is nothing but the result of  the new bosonic symmetry studied in this paper acting on the gauge field.  Although we presented the details of the Abelian theory above for the sake of transparency, the supersymmetry transformations readily generalize to the non-Abelian  case\cite{aa-msusy1}. Thus, the bosonic symmetry studied here can also be regarded as a consequence of supersymmetry, in the sense that the massive supersymmetry algebra contains this bosonic  symmetry. 

We take this opportunity to correct an interpretational  issue related to the closure of the algebra(\ref{closure}). From the explicit expression of the current $J_\mu  = A_\mu -  \partial_\mu\frac{1}{\partial_\alpha \partial^\alpha}\partial^\nu A_\nu$ in the Abelian case, it is  tempting to think of the second term in (\ref{closure}) as a combination of a spacetime rotation and the composition of a rotation and gauge transformation of $A_\mu$. Although the supersymmetry transformations given in \cite{aa-msusy1} are valid, this interpretation of the underlying massive algebra suggested in that paper is inconsistent. The issue is that one cannot have the commutator of the supercharges ($\mathcal{Q}$) close on spacetime rotation generators. For instance, the Jacobi identity 
\[
[\bar\epsilon \mathcal{Q},[\bar\beta \mathcal{Q},P]]+ [\bar\beta \mathcal{Q},[P,\bar\epsilon \mathcal{Q}]] + [P,[\bar\epsilon \mathcal{Q},\bar\beta \mathcal{Q}]] = 0
\]
cannot be consistently satisfied if the commutators of the supercharges involve a term proportional to the spacetime rotation generators since the first two terms above will vanish while the last third term on the $r.h.s$ will be generically non-zero. However, regarding the deformation of the transformation transformation $\delta A_\mu =  2im\epsilon_{\mu \nu \lambda}J^\mu (\bar \omega \sigma ^\lambda \rho)$ as a symmetry that involves as a non-local field redefinition of the gauge potential (and hence commutes with the momentum generator $P$) avoids this interpretational problem completely. Thus not only is the new current-field symmetry implied by considerations of supersymmetry it is also necessary for the consistency of the mass-deformed supersymmetry algebra. 
\section{Noether current}

Continuous symmetries of Lagrangians imply, according to Noether's theorem, the existence of corresponding conserved currents.  The charges associated with the currents generate the symmetries.  Textbooks tend to focus on the case when the highest number of derivatives in the Lagrangian is 2, a situation in which the construction of the Noether current is particularly straightforward.  The textbook method can easily be extended to theories with any number of derivatives capped by some fixed number.  In the case at hand, however, we are dealing with the an infinite number of derivatives due to the presence of a non-local term in the Lagrangian.  In general, it is difficult to find a method for deriving the Noether current.  Nevertheless, in the present situation, it is relatively simple to make a plausible guess for the Noether current.

Without any pretense of rigor in the "derivation", we will construct a conserved current.  If we do not worry about operator ordering and the commutation properties of $J$ and $A$, (\ref{symm}) is generated by:
\be
v^\lambda Q_\lambda = v^\lambda \int d^2x \epsilon_{\lambda \mu \nu}\Pi^{a\mu}(x) J^{a\nu}(x)
\ee
where $\Pi^{a\mu}(x)$ is the conjugate momentum of $A^a_\mu(x)$:
\be
[A^a_\mu(x), \Pi^{b\nu}(y)]=i\delta^{ab}\delta^\nu_\mu \delta^{(2)}(x-y)
\ee
Instead of attempting to quantize the full non-local theory and computing the conjugate momentum of the gauge field, we will simply consider the simplest case where $m=0$ and try to accommodate the $m\neq 0$ case as needed.  Since in the pure Yang-Mills case, the canonical momentum of $A_0$ vanishes, we will proceed by setting it to zero:
\be
v^\lambda Q_\lambda = v^\lambda \int d^2x \epsilon_{\lambda i \nu}E^{ai}(x) J^{a\nu}(x)
\ee
Since $Q_\lambda$ should be the space integral of the time component of the Noether current, $Q_\lambda = \int d^2 M_{0\lambda}$ where $M_{0\lambda}$ is the $0$ component of a current, then one might make the following naive guess for the Noether current:
\be
M_{\sigma \mu} = g_{\sigma\delta}\epsilon_{\mu\nu\rho}F^{a\delta\nu}J^{a\rho}
\ee
However, this current is not conserved
\be
\partial^{\sigma}M_{\sigma\mu} =\epsilon_{\mu\nu\tau}(m^2J^{a\nu}J^{a\rho} +  F^{a\delta\nu}D_\delta J^{a\tau}) = \epsilon_{\mu\nu\tau}F^{a\delta\nu}D_\delta J^{a\tau}
\ee
We made use of the equation of motion:$D_\mu F^{\mu\nu} = m^2J^\nu$ above.  By using (\ref{inv}), the above expression can be put into the following convenient form:
\be
\partial^{\sigma}M_{\sigma\mu}  =  \frac{1}{2}\epsilon_{\rho\nu\tau}F^{a\rho\nu}D_\mu J^{a\tau}
\ee
Given the variation of the action (\ref{varact}), it is natural to consider a modified current:
\be
L_{\sigma\mu} = g_{\sigma\delta}\epsilon_{\mu\nu\tau} F^{a\delta\nu}J^{a\tau} -\frac{1}{2}\epsilon_{\rho\nu\sigma} F^{a\rho\nu}J^{a}_{\mu}
\ee
In the previous section we proved (\ref{varact}) that:
\be
\partial^\rho(\epsilon_{\mu\nu\rho}F^{\mu\nu a}J_\sigma^a) = \epsilon_{\mu\nu\rho}F^{a\mu\nu}(D_\sigma J^\rho)^a
\ee
Using the above equation, 
it is straightforward to prove the following identities:
\bea
\partial^\sigma L_{\sigma\rho} = 0 \nonumber \\
\partial^\rho L_{\sigma\rho} = 0 \nonumber \\
L_\sigma^\sigma = \frac{1}{2}\epsilon_{\rho\nu\tau} F^{a\rho\nu}J^{a\tau}
\eea

The current $L_{\mu\nu}$ is conserved both in the $m=0$ and $m\neq 0$ cases.
The charges computed from the above currents:
\be
Q_\mu = \int d^2x L_{0\mu} =  \int d^2x(-\epsilon_{\mu i \tau} F^{a0i}J^{a\tau} -F^{a12}J^{a}_{\mu})
\ee
differ from the ones we started with in the $m=0$ case but still generate the transformation (\ref{symm}) if $[A_\mu, J_\nu]=0$.  In the Abelian case, where $J$ has a particularly simple expression, one can check that the current and gauge field do indeed commute.   

\section{The symmetry algebra in the Abelian case}
The Abelian case is considerably simpler than the general one.  In this case, we are able to provide closed form expressions for $J$ and compute the algebra of charges, $[Q_\mu, Q_\nu]$, by evaluating the  commutator under (\ref{symm}):
\be
(\delta'\delta - \delta\delta')A_\mu = \delta\theta^\nu \delta\theta'^\kappa (\epsilon_{\kappa\nu}^{\;\;\;\;\lambda}\epsilon_{\lambda\mu\rho}J^\rho +  \epsilon_{\kappa\nu}^{\;\;\;\;\lambda}\epsilon_{\mu\eta\rho}\partial_\lambda\frac{1}{\partial_\alpha\partial^\alpha}F^{\eta\rho}) \label{1comm}
\ee
The first term on the right hand side is a variation of the same form as (\ref{symm}).  If the second term on the right hand side were absent, the algebra of charges would be SO(2,1).  However, this is not the case, and since the second term is not of the form (\ref{symm}), the symmetry algebra does not close on the generators of (\ref{symm}) alone.  It remains to see whether the algebra can be made to close by adding a finite number of new generators.  

The second term on the right hand side of (\ref{1comm}) suggests a new symmetry of the action:
\be
\delta_2 A_\mu = \delta\psi^\sigma \partial_\sigma \epsilon_{\mu\nu\kappa}\frac{1}{\partial_\alpha\partial^\alpha}F^{\nu\kappa} \label{symm2}
\ee
One can check that this is indeed a symmetry of the mass deformed action for all values of $m$ ( including $m=0$).  
The commutator of (\ref {symm2}) with itself vanishes:
\bea
(\delta_2\delta_2' - \delta_2'\delta_2)A_\mu = 0 \label{22comm}
\eea
We consider next the commutator of the symmetries (\ref{symm}) and (\ref{symm2}):
\bea
(\delta_1\delta_2 - \delta_2\delta_1)A_\mu = 2\delta\theta^\sigma\delta\psi^\rho\partial_\mu\partial_\rho\frac{1}{\partial_\alpha\partial^\alpha}J_\sigma \label{12comm}
\eea
For clarity, we have introduced the notation $\delta_1$ as the variation that we called $\delta$ in (\ref{symm}).
The right hand side corresponds to transformations:
\be
\delta_3A_\mu = \gamma^{\rho\sigma}\partial_\mu\partial_\rho\frac{1}{\partial_\alpha\partial^\alpha}J_\sigma \label{symm3}
\ee
where, $\gamma^{\rho\sigma}$ is a traceless ($\gamma^{\rho\sigma}g_{\rho\sigma}=0$) matrix of constant parameters.  Because of the $\partial_\mu$ appearing in (\ref{symm3}), the transformation (\ref{symm3}) is a gauge transformation and, therefore, both $F_{\mu\nu}$ and $J_\mu$ are invariant under (\ref{symm3}) leaving the action invariant, trivially.   It is useful to split $\delta_{3}$ into two variations according to the symmetry properties of $\gamma^{\rho\sigma}$:
\bea
&&\delta_3 = \delta_{3A} + \delta_{3S}  \\ 
&& \delta_{3A}A_\mu = \alpha^{\rho\sigma}\partial_\mu\frac{1}{\partial_\alpha\partial^\alpha}F_{\sigma\rho} \\
&&\delta_{3S}A_\mu = \beta^{\rho\sigma}\partial_\mu(\partial_\rho\frac{1}{\partial_\alpha\partial^\alpha}J_\sigma +  \partial_\sigma\frac{1}{\partial_\alpha\partial^\alpha}J_\rho) 
\eea
where 
\be
 \alpha^{\rho\sigma} = - \alpha^{\sigma\rho} \; , \;  \beta^{\rho\sigma} = \beta^{\sigma\rho} \; ,\; \beta^{\rho\sigma}g_{\rho\sigma} = 0.
\ee

Defining generators for the transformations (\ref{symm}), (\ref{symm2}), and (\ref{symm3}) as:
\bea
\delta_1 A_\mu &=& [i\delta\theta^\nu Q_{\nu}^{(1)}, A_\mu] \\
\delta_2 A_\mu &=& [i\delta\psi^\nu Q_{\nu}^{(2)}, A_\mu] \\
\delta_{3A} A_\mu &=& [i\delta\alpha^{\rho\nu} Q_{\rho\nu}^{(3A)}, A_\mu] \\
\delta_{3S} A_\mu &=& [i\delta\beta^{\rho\nu} Q_{\rho\nu}^{(3S)}, A_\mu] 
\eea
The relations (\ref{1comm}), (\ref{12comm}) and (\ref{22comm}) can be expressed as:
\bea
&&[Q_{\mu}^{(1)}, Q_{\nu}^{(1)}] = i\epsilon_{\mu\nu}^{\;\;\;\;\lambda}(Q_{\lambda}^{(1)} + Q_{\lambda}^{(2)}) \\
&&[Q_{\mu}^{(2)}, Q_{\nu}^{(2)}] = 0 \\
&&[Q_{\mu}^{(1)}, Q_{\nu}^{(2)}] = i(Q_{\mu\nu}^{(3A)} + Q_{\mu\nu}^{(3S)}) \\
\eea

In addition to the above commutation relations, we should compute commutators of $Q_{\rho\nu}^{(3A)}, Q_{\rho\nu}^{(3S)}$ with all other generators, for completeness.  However, since $Q_{\rho\nu}^{(3A)}, Q_{\rho\nu}^{(3S)}$ generate gauge transformations on $A_\mu$, their commutators with all other generators are also gauge transformations.  This can be seen as follows
\bea
\delta_3\delta_iA_\mu = 0 
\eea 
since $\delta_i A_\mu$ is gauge invariant for any $i$, but
\be
\delta_i\delta_3A_\mu = \gamma^{\rho\sigma}\partial_\mu\partial_\rho\frac{1}{\partial_\alpha\partial^\alpha}\delta_iJ_\sigma 
\ee
Therefore, 
\be
(\delta_i\delta_3 - \delta_3\delta_i)A_\mu = \partial_\mu(\gamma^{\rho\sigma}\partial_\rho\frac{1}{\partial_\alpha\partial^\alpha}\delta_iJ_\sigma)
\ee
is a gauge transformation and the algebra generated by $Q_{\nu}^{(1)}$ and $Q_{\nu}^{(2)}$ closes up to gauge transformations.  In particular, 
\bea
[Q_{\mu}^{(1)}+ Q_{\mu}^{(2)}, Q_{\nu}^{(1)} + Q_{\nu}^{(2)}] = i\epsilon_{\mu\nu}^{\;\;\;\;\lambda}(Q_{\lambda}^{(1)} + Q_{\lambda}^{(2)}) +  2iQ_{\mu\nu}^{(3A)}
\eea
is an SO(2,1) sub-aglebra up to gauge transformations (since $Q_{\mu\nu}^{(3A)}$ generates gauge transformations).  This $SO(2,1)$ symmetry ($SO(3)$ in the Euclidean case) is distinct from Lorentz transformations.
In general, for the non-Abelian, as well as the Abelian case, iterated commutators of the current-field symmetry transformations generate new non-local transformations. It would obviously be extremely interesting to try and understand the algebraic structure behind the  symmetries obtained by these iterated commutators.

\section{Concluding remarks}
 Several questions that require further attention are naturally prompted by this study.  The most obvious question to ask is whether or not there are solutions to (\ref{fund-constraints}) in three spacetime dimensions. Making a concrete statement about the existence of solutions to these equations would be interesting in its own right. If these constraints can be solved, establishing the  putative non-local symmetry in three dimensions, a pressing issue would be to understand the precise structure of this symmetry.  As shown in the last section, even the Abelian case has a lot of structure.  We found that the algebra of symmetries (\ref{symm}) does not close on itself and it is necessary to extend (\ref{symm}) to include a new set of symmetry generators (\ref{symm2}).  The algebraic structure has some intriguing features and it would be interesting to understand the corresponding story in the general non-Abelian case.

Also, as pointed out earlier in the paper, a better understanding of the interplay between the current-field symmetry  and supersymmetry in three dimensions is needed for understanding various issues related to the interplay between supersymmetry and mass-gaps in the spectra of supersymmetric gauge theories in $D=2+1$. We hope to address these issues in the near future.

Another interesting set of questions has to do with whether or not there is a Hamiltonian counterpart to this new symmetry. Since the symmetry discussed in this paper involves transformations that are non-local both in space and time, generalizing this discussion to a Hamiltonian framework of a non-Abelian theory is not straightforward. Whether or not this symmetry plays a role in the gauge invariant Hamiltonian framework of the three dimensional gauge theory \cite{KKN-Papers} is obviously an open  and interesting question.

It is also natural to ask, if the existence of solutions of (\ref{fund-constraints}) violates the Coleman-Mandula theorem\cite{cm}.  The Coleman-Mandula theorem requires the symmetry generators of  interacting Lorentz invariant  theories to be Lorentz scalars, which the symmetry generators are obviously not. The two cases where we have been able to construct $J_\mu $ explicitly in this paper - an interacting theory in $1+1$ dimensions, and a free (abelian) theory in $2+1$ dimensions - are both exempt from the domain of applicability of the theorem. A non-abelian solution to  (\ref{fund-constraints})  in three spacetime dimensions might, at least naively, appear to be at odds with the  Coleman-Mandula theorem.  However, given that we expect such a current to be both non-local in space and time it might also be exempt from the Coleman-Mandula theorem, since an implicit locality of the symmetry generators is assumed in the derivation of the theorem. 

Finally, on a speculative note, we comment on the  iterated commutators of the non-local current-field symmetry transformation which were shown in the last section to generate new non-local symmetries. It is tempting to note that  the existence of higher conservation laws generated by iterated commutators of a non-local symmetry operation, is reminiscent of Yangian symmetries that integrable systems, such as the Heisenberg spin chain, often possess. In the context of a higher dimensional quantum field theory Yangian symmetries have recently been shown to play a crucial role in rendering the Hamiltonian of $\mathcal{N}=4$ super Yang-Mills on $\mathcal{R}\times S^3$ integrable (for recent reviews and older pedagogical introductions to the subject see \cite{yang-review1, yang-review2, aayang,wittenyang}). It would be natural  to explore if the spacetime $SO(2,1)$ Lorentz symmetry fuses with the non-local symmetry transformation laws given here to form a Hopf algebraic structure, such as a Yangian. Perhaps a fruitful starting point for understanding this issue is the $D=1+1$ version of the current field symmetry discussed earlier in the paper. Since the pure glue theory in two dimensions is an integrable system\cite{apmin}, as is the Wess-Zumino-Witten model,  understanding how the current-field symmetry is embedded in the infinite tower of conserved charges of these theories might help us understand its algebraic properties in the three dimensional case as well.  
\\
\\
{\bf Acknowledgements:} We are grateful to V.P.Nair for many illuminating discussions and his constant encouragement.
  
\end{document}